\documentclass[draft]{agujournal2019}
\usepackage{url} 
\usepackage{lineno}
\usepackage[inline]{trackchanges} 
\usepackage{soul}
\draftfalse

\journalname{Geophysical Research Letters}

\begin{document}

%
%

\title{Reply to comment on “An Active Plume Eruption on Europa During Galileo Flyby E26 as Indicated by Energetic Proton Depletions”}

%
%




\authors{H.L.F. Huybrighs\affil{1}, E. Roussos\affil{2}, A. Blöcker\affil{3, 4}, N. Krupp\affil{2}, Y. Futaana\affil{5}, S. Barabash\affil{5}, L.Z. Hadid\affil{6}, M.K.G. Holmberg\affil{1}, O. Witasse\affil{1}}


\affiliation{1}{ESA/ESTEC, Noordwijk, The Netherlands}
\affiliation{2}{Max Planck Institute for Solar System Research, Göttingen, Germany}
\affiliation{3}{KTH, Royal Institute of Technology, Stockholm, Sweden}
\affiliation{4}{Department of Earth and Environmental Sciences, Ludwig Maximilian University of Munich, Munich, Germany}
\affiliation{5}{IRF, Swedish Institute of Space Physics, Kiruna, Sweden}
\affiliation{6}{LPP, CNRS, École polytechnique, Sorbonne Université, Observatoire de Paris, Université Paris-Saclay, PSL Research University, Palaiseau, France}





\correspondingauthor{H.L.F. Huybrighs}{hans.huybrighs@esa.int}




\begin{keypoints}
\item EPD artefact does not affect conclusion that perturbed fields and atmospheric charge exchange  drive energetic proton losses at Europa
\item Energetic proton measurements could be used to infer the presence of plumes with density profiles consistent with the literature
\item The artefact reported by \citeA{Jia2021} prevents a definite conclusion on the presence of a plume during E26 from EPD data
\end{keypoints}

%
%

%
%


\begin{abstract}
In \citeA{Huybrighs2020} we investigated energetic proton depletions along Galileo’s Europa flyby E26. Based on a particle tracing analysis we proposed that depletions are caused by perturbed electrogmagnetic fields combined with atmospheric charge exchange and possible plumes. One depletion feature identified as a plume signature was shown to be an artefact \cite{Jia2021}. 
Despite that, here we emphasize that \citeA{Huybrighs2020} demonstrates that plumes can cause proton depletions and that these features should be sought after. Furthermore, the conclusions on the importance of perturbed electromagnetic fields and atmospheric charge exchange on the depletions are unaffected.
We suggest that the artefact’s cause is a mistagging of protons as heavier ions by EPD. The artefact prevents us from confirming or excluding that there is a plume associated depletion. We also address comments on the MHD simulations and demonstrate that 540-1040 keV losses are not necessarily inconsistent with 115-244 keV losses by plume associated charge exchange.
\end{abstract}

\section*{Plain Language Summary}
In \citeA{Huybrighs2020} we identified why fast protons were disappearing during Europa flyby E26 by Galileo. Beyond impacting on the surface we identified several contributing factors: Firstly, perturbed electromagnetic fields resulting from the interaction of Europa's atmosphere with the magnetospheric plasma, which deflect the protons. Secondly, atmospheric charge exchange. We also showed that a water plume eruption could cause a region in which disappearances occur due to a combination of charge exchange and magnetic deflections. We identified a 20s decrease of protons as evidence of such a plume. However, an artefact in the data reported by \citeA{Jia2021} coincides with this 20s moment and prevents us from reaching a conclusion on the occurrence of a plume associated depletion. We emphasize that our conclusions on the importance of perturbed fields and charge exchange are unaffected, as the artefact only affects a short segment of the data we analysed. 
Furthermore, our results demonstrate that plumes can cause proton depletions and that these features should be sought after in the data.  

%
%

%


%
%
%
%

\section{Unaffected implications of Huybrighs et al., 2020}
The title of \citeA{Huybrighs2020} only highlighted the potentially important result regarding the plume feature. Since this title is unavoidably retained across the different commentaries, it could give the wrong impression that our original study is invalid. However, the majority of the results in \citeA{Huybrighs2020} are not affected by the data artefact identified by \citeA{Jia2021}. In addition, the simulation results of plume-induced proton losses still hold, indicating the feasibility of tracking neutral gas inhomogeneities, such as plumes, with energetic proton data. 
The artefact is shown in Figure \ref{fig_tp_th} as periodically occurring once every $\sim$120 seconds, corresponding to the intervals with the EPD instrument on motor position seven. Each instance of the data artefact lasts one spin ($\sim20$ seconds), which is significantly shorter than the period over which interactions with the energetic protons and Europa can be identified (approximately 4 minutes). Thus the majority of the depletion features we investigated are not affected. Specifically:
\begin{itemize}
    \item The measured proton depletion along the trajectory in effect represents a 'tomography' of the perturbed fields and atmosphere-ion interaction of Europa, as the protons scan the interaction region by travelling across different regions of it.  
    The perturbed fields act to deflect the protons away from the surface and decrease the contribution of impacting particles to the total loss. Whereas previous studies, e.g. \citeA{Cassidy2013,Breer2019}, concentrated on the modified surface precipitation patterns of energetic ions and electrons through simulations \citeA{Huybrighs2020} shows the impact of the electromagnetic field perturbations on directional energetic particle measurements at higher altitudes ($>350$ km).
    \item Our results provide evidence for the role of atmospheric charge exchange in the formation of energetic proton depletions in the range 115-244 keV.
    \item Plumes can cause localised energetic proton depletions through charge exchange (115-244 keV) and/or their associated magnetic field perturbations (115keV-1.04 MeV), thereby emphasizing the importance of energetic ion measurements in close Europa flybys.
\end{itemize}

\section{EPD data artefact and implications for a plume detection during E26}
We use this opportunity to elaborate on the origin of the poorly documented artefact, for reference in future investigations relying on EPD observations. The data artefact identified by \citeA{Jia2021} was briefly reported earlier in \citeA{Kollmann2016}, but went unnoticed in \citeA{Huybrighs2020} since a different archived dataset was used and the comments of \citeA{Kollmann2016} on the artefact did not propagate in the documentation of the archived dataset. Furthermore, in the complete time frame considered for \citeA{Huybrighs2020} two instances of motor seven occur. However, at the first motor seven position curiously no artefact occurs.

\begin{figure}
\includegraphics[width=1.0\textwidth]{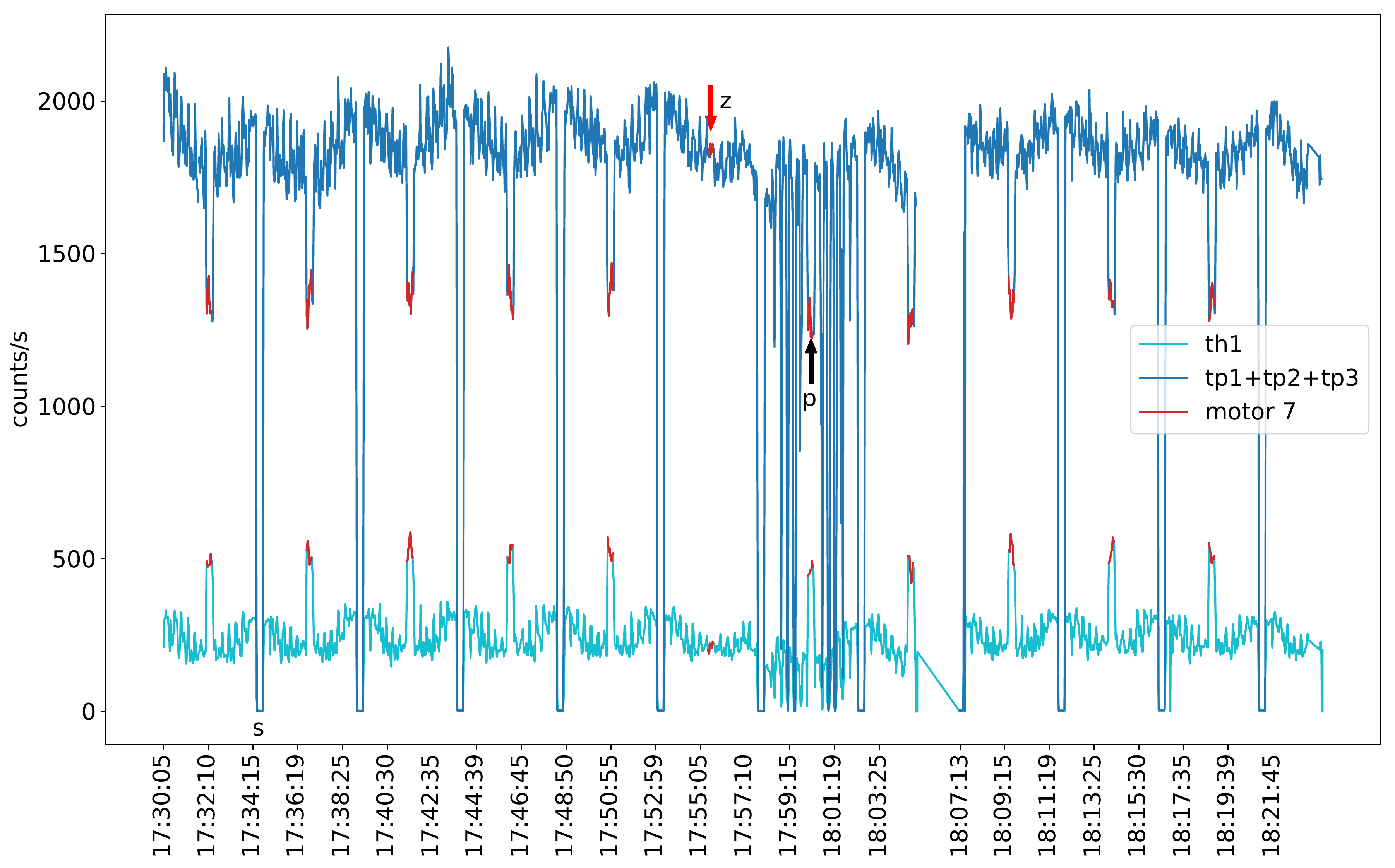}
\caption{Sum of the TP channels (dark blue) versus the TH1 channel (light blue). Note that at motor seven (marked in red), the TP channels are anti-correlated to TH1.
'p' refers to the depletion attributed to a plume in \citeA{Huybrighs2020}. 'z' is the motor seven position from \citeA{Huybrighs2020} unaffected by the artefact.
The depletions to zero (first one marked 's') correspond to times when the detector passes behind the foreground shield.}
\label{fig_tp_th}
\end{figure}

We found that the depletions at motor position seven are anti-correlated to the TH1 channel (see Figure \ref{fig_tp_th}), which is designed to record ions heavier than sulphur. In Figure \ref{fig_tp_th} a larger time range is shown than in \citeA{Huybrighs2020}. Increases in the counts in TH1 at motor seven (marked in red) correspond to decreases at motor seven in the count rate of the TP channels, which measure protons. At the first motor seven position in the timeframe we considered in \citeA{Huybrighs2020} (marked 'z' in Figure \ref{fig_tp_th}), no increase in TH1 is present, which could explain why TP measurements are not affected. 
TH1/TP1 measurements free of the sector seven artefacts occur without any obvious pattern. Their occurrence just before the E26 closest approach (CA) was a misleading coincidence that prevented us from identifying the issue reported in \citeA{Jia2021}. The motor seven position corresponding to 'p', the depletion originally attributed to a plume in \citeA{Huybrighs2020}, does coincide with an increase in TH1. Furthermore we have identified that in the channels TP2, TP3, TO1, TO2, TO3, TO4 and TS3 sector seven is also anti-correlated to TH1. We thus find that there are too many unknowns to formulate a relation between the channels that can be used to compensate for the artefact. 

The core of the problem appears to be with the counts of the start detector of EPD/CMS Time of Flight (TOF) system showing a previously unnoticed increase at sector seven correlated with sector seven of TH1. We hypothesise that excess noise-driven counts of the START detector of the CMS's TOF system correlate with stop counts of ion events, leading to a wrong TOF estimate for those events and their eventual mistagging as part of the TH1 channel.

Correcting the artificial TP sector seven dropout gives satisfactory results when the excess counts of TH1 at sector seven are added to the reduced counts recorded by the sum of the TP channels. Still, even if a minor dropout in TP1 remains at 'p', it is still smaller than the uncertainties associated with an empirical correction. Should the residual dropout be considered as real, it would put an upper limit of the plume surface density of $2.5\times10^8$cm$^{-3}$ based on Figure 2 in \citeA{Huybrighs2020}. We conclude that due to the uncertain nature of the artefact and its precise contribution to TP1 we can neither confirm nor exclude that a plume associated depletion is present in the TP1 measurement.

\section{MHD simulations}
\citeA{Bloecker2016} has previously established that the fields during flyby E26 are not consistent with a homogeneous atmosphere. In addition \citeA{Arnold2019} has identified features in the magnetic field that are consistent with a plume. We used the plume location suggested by \citeA{Arnold2019} as input for the MHD simulations in \citeA{Huybrighs2020}.

\begin{figure}
\includegraphics[width=1.0\textwidth]{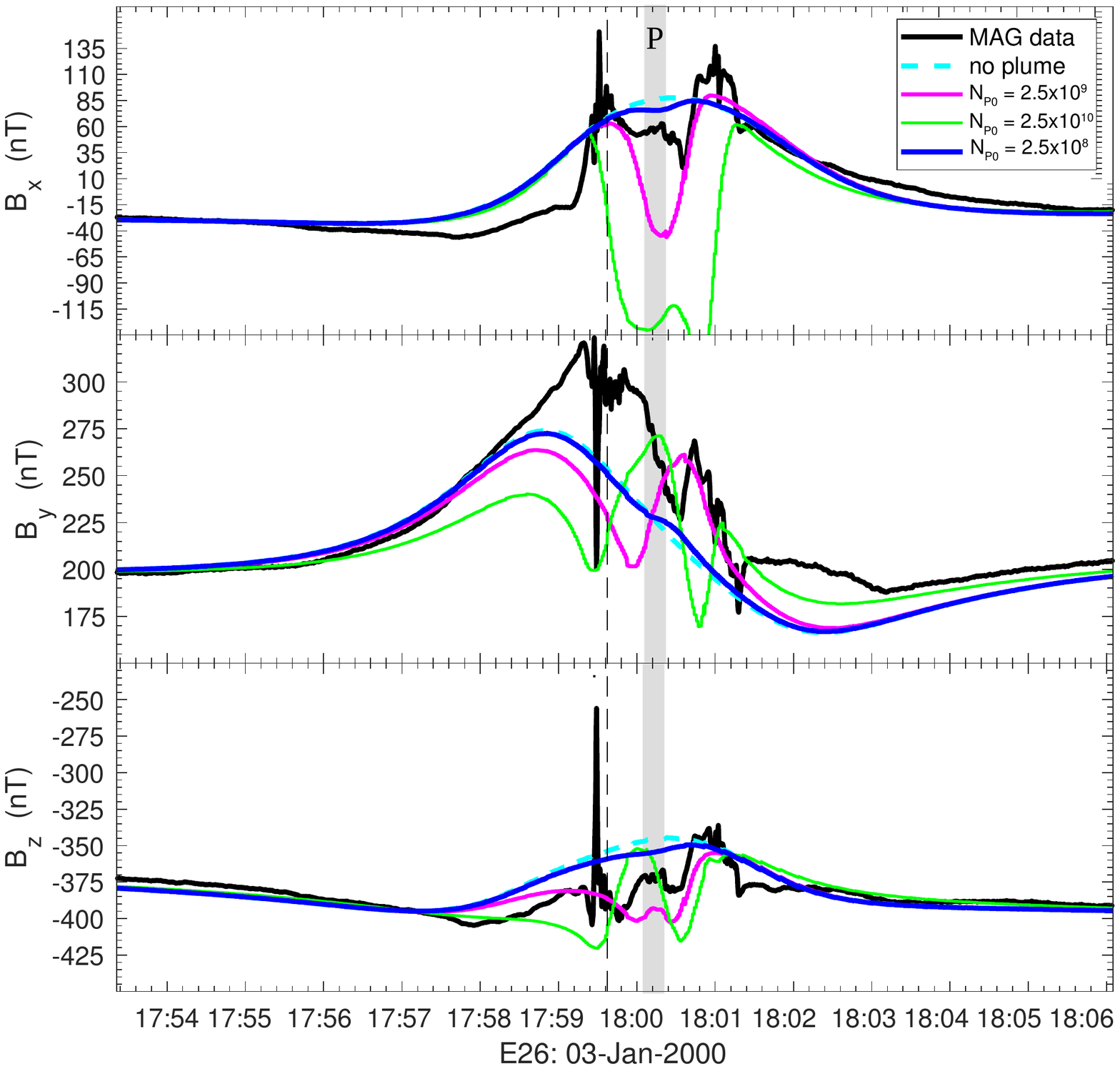}
\caption{E26 flyby MAG data and simulations for different plume densities $N_{P0}$ in $cm^{-3}$ (corresponding to those in Figure \ref{fig_e26_tp3_zoom}). The vertical dashed line represents the closest approach.}
\label{fig_mag_galileo_E26}
\end{figure}

\citeA{Jia2021} judges the effect of the plume on the fields on time frame 'p'. Here we want to point out that the effect of a plume is not confined to this $\sim$20s time frame, which is illustrated by Figure \ref{fig_mag_galileo_E26}. We find that the MHD simulation without a plume does not reproduce the bi-lobal signature in Bx and Bz and the bump-like feature in By after the closest approach, while the strong inhomogeneity in neutral densities induced by a plume tends to cause these aspects. 
We consider that the situation with a homogeneous atmosphere is not a good reproduction of the data. While the inclusion of the \citeA{Arnold2019} plume in our MHD simulations still leaves several features unexplained, especially in By, the trends in the data/simulation comparison are suggestive of an atmosphere with a local inhomogenity, which is why we considered  plume-driven depletions in EPD as a reasonable scenario to explore.
Follow up studies should strive towards identifying a case that fits both the EPD and MAG data. However, due to the large amount of unknown parameters (atmospheric properties, plume properties) we are unable to identify such a case here, despite being suggestive that plumes are indeed viable. 

\section{Proton depletions at 0.54-1.04 MeV due to a plume}
\begin{figure}
\includegraphics[width=1.0\textwidth]{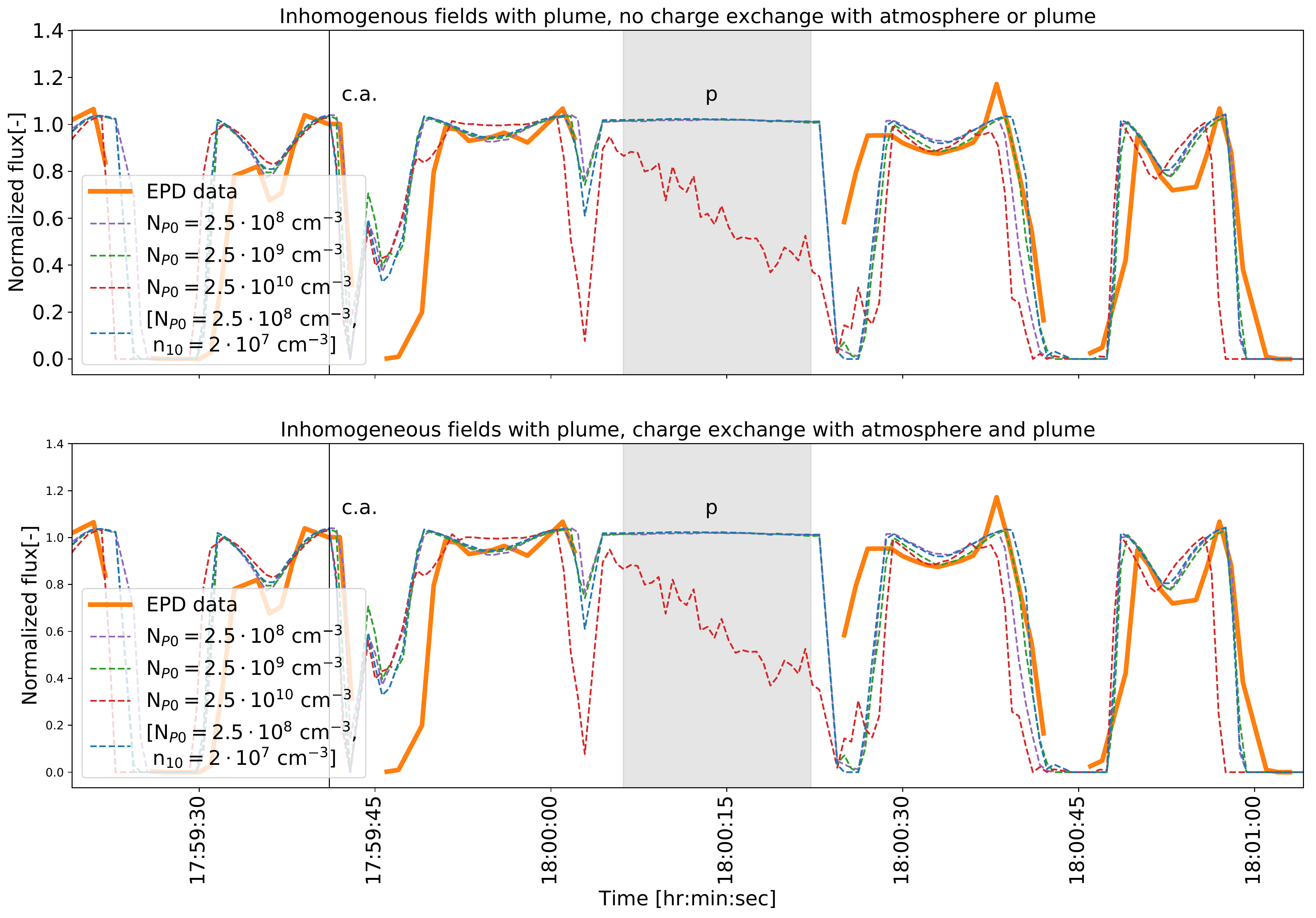}
\caption{E26 flyby, TP3 channel (0.54-1.04 MeV). The orange curve shows the data. The dashed curves show different simulation cases in which the plume density $N_{P0}$ has been varied. The solid black line indicates the closest approach (c.a.). The label 'p' indicates the potential plume feature. Note that the TP3 data in 'p' has been left out, as it is affected by the artefact.}
\label{fig_e26_tp3_zoom}
\end{figure}

\citeA{Jia2021} elaborate that a depletion occurring at both TP3 (0.54-1.04 MeV) and TP1 (114-244 keV) are inconsistent with a plume. Charge exchange is negligible in TP3 due to the orders of magnitude reduction in the charge exchange cross section compared to TP1. We should state here that in light of \citeA{Jia2021}, we consider that the TP3 dropout at 'p' is likely an artefact of a similar nature as that of TP1. Regardless, here we want to emphasize that a depletion that occurs both at the high and low energy channels is not necessarily inconsistent with the presence of a charge exchange driven depletion by a plume or atmosphere. 
In Figure \ref{fig_e26_tp3_zoom} we show the simulations corresponding to the higher energy channel TP3, analogous to Figure 2 in \citeA{Huybrighs2020} for the TP1 channel. A plume with the highest density ($N_{P0}$) considered in \citeA{Huybrighs2020} causes depletions in the highest energy channel as well as the low energy channel. Though this case is  inconsistent with the MAG and EPD data, it does demonstrate that it is in principle possible for a plume to cause a depletion at both the low energy channel TP1 and TP3. In TP1 depletions occur both due to charge exchange and field perturbations, at the TP3 channel only due to field perturbations because of the strongly reduced charge exchange cross section. 

\section{Conclusion}
The attributions of energetic proton losses during flyby E26 to charge exchange and electromagnetic field perturbations by \citeA{Huybrighs2020} still hold, with the exception of a short-duration dropout originally attributed to a plume, which is due to a data artefact as pointed out by \citeA{Jia2021}. 
Because of the artefact we can neither confirm nor exclude that a plume associated depletion is present in the TP1 data. Despite that, the simulation results of \citeA{Huybrighs2020} hold in the sense that plume and atmosphere driven depletions of energetic ions are feasible and should be sought for in existing or future datasets.

\acknowledgments
We acknowledge Xianzhe Jia, Margaret G. Kivelson and Christopher Paranicas for their effort in analysing our manuscript \citeA{Huybrighs2020}. 
EPD data are available through NASA's Planetary Data System (PDS) https://pds-ppi.igpp.ucla.edu/mission/Galileo/GO/EPD.


%
%

\bibliography{database}

%
%
%
%
%

\end{document}